\documentclass[]{IAC_style}
\PassOptionsToPackage{hyphens}{url}
\usepackage[hidelinks]{hyperref}
\begin{document}

\IACpaperyear{2025} % format yyyy
\IACpapernumber{IAC-25-A6.7.8.x100237} % full paper id
\IAClocation{Sydney, Australia} % used in the header
\IACdate{29 Sep-3 Oct 2025} % used in the header
%76th International Astronautical Congress (IAC 2025), Sydney, Australia, 29 Sep-3 Oct 2025. 
% Which copyright? If B, put the copyright holder.
%\IACcopyrightA{}
\IACcopyrightB{the International Astronautical Federation (IAF)}
%IAC-25,A6,7,8,x100237
\title{Analyzing Data Quality and Decay in Mega-Constellations: A Physics-Informed Machine Learning Approach}

%\IACauthor{Author name}{corresponding affiliation: nr. of corresponding affiliation}{is corresponding author? 0-1}
\IACauthor{Katarina Dyreby$^{\orcidlink{0000-0000-0000-0000}}$}{1}{0}
\IACauthor{Francisco Caldas$^{\orcidlink{0000-0000-0000-0000}}$}{2}{0}
\IACauthor{Cláudia Soares$^{\orcidlink{0000-0000-0000-0000}}$}{3}{0}

% Input affiliations here, order is relevant
\IACauthoraffiliation{FCT-UNL, Portugal,\normalfont{~\authormail{k.dyreby@campus.fct.unl.pt}}}
\IACauthoraffiliation{FCT-UNL, Portugal\normalfont{~\authormail{f.caldas@campus.fct.unl.pt
}}}
\IACauthoraffiliation{FCT-UNL, Portugal\normalfont{~\authormail{claudia.soares@fct.unl.pt}}}

%\abstract{In the era of mega-constellations, the need for accurate and publicly available information has be-
%come fundamental for satellite operators to guarantee the safety of spacecrafts and the Low-Earth Orbit
%(LEO) space environment. This study critically evaluates the accuracy and reliability of publicly available
%ephemeris data for a LEO mega-constellation – Starlink.
%The goal of this work is twofold: (i) compare and analyze the quality of the data against high-
%precision numerical propagation. (ii) Leverage Physics-Informed Machine Learning to extract relevant
%satellite quantities, such as mass and drag coefficients, and compare them against known information,
%namely, during the decay process.
%By analyzing two months of real orbital data for approximately 1500 Starlink satellites, we identify
%discrepancies between high precision numerical algorithms and the published ephemerides, recognizing
%the use of simplified dynamics at fixed thresholds, planned maneuvers, and limitations in uncertainty
%propagations.
%Furthermore, we compare data obtained from multiple sources to track and analyze deorbiting satellites
%over the same period. Empirically, we extract the acceleration profile of satellites during deorbiting and
%provide insights relating to the effects of non-conservative forces during reentry.
%Through this in-depth analysis, we highlight potential limitations in publicly available data for accurate
%and robust Space Situational Awareness (SSA), and importantly, we propose a data-driven model of
%satellite decay in mega-constellations.}

\abstract{In the era of mega-constellations, the need for accurate and publicly available information has become fundamental for satellite operators to guarantee the safety of spacecrafts and the Low Earth Orbit (LEO) space environment. This study critically evaluates the accuracy and reliability of publicly available ephemeris data for a LEO mega-constellation – Starlink.
The goal of this work is twofold: (i) compare and analyze the quality of the data against high-precision numerical propagation. (ii) Leverage Physics-Informed Machine Learning to extract relevant satellite quantities, such as non-conservative forces, during the decay process.
By analyzing two months of real orbital data for approximately 1500 Starlink satellites, we identify discrepancies between high precision numerical algorithms and the published ephemerides, recognizing the use of simplified dynamics at fixed thresholds, planned maneuvers, and limitations in uncertainty propagations. 
Furthermore, we compare data obtained from multiple sources to track and analyze deorbiting satellites over the same period. Empirically, we extract the acceleration profile of satellites during deorbiting and provide insights relating to the effects of non-conservative forces during reentry. For non-deorbiting satellites, the position Root Mean Square Error (RMSE) was approximately 300 m, while for deorbiting satellites it increased to about 600 m.
Through this in-depth analysis, we highlight potential limitations in publicly available data for accurate and robust Space Situational Awareness (SSA), and importantly, we propose a data-driven model of satellite decay in mega-constellations.}

\maketitle
\thispagestyle{fancy} % resets proper header/footer

%\section*{Nomenclature}

\section{Introduction}
\vspace{0.3cm}

\noindent As the number of active satellites in Low Earth Orbit (LEO) continues to grow, ensuring their safe operation has become a complex challenge. Accurate trajectory prediction and collision avoidance are now essential, as overcrowding in LEO has significantly raised the likelihood of orbital collisions ~\cite{ESA2025}. Such events not only threaten the functionality of space assets but also contribute to the accumulation of debris, increasing the risk of chain reaction scenarios like the Kessler syndrome~\cite{kessler1978collision}. This theoretical scenario hypothesizes that cascading collisions could cause an exponential increase in debris, eventually saturating orbits and making them unusable.

Large scale satellite constellations, such as Starlink, have further intensified these concerns by increasing the density of satellites at an unprecedented rate. To manage the operational risks associated with these mega-constellations, continuous and precise tracking of active satellites is more important than ever before. Starlink, operated by SpaceX, is currently the largest satellite mega constellation, with over $9058$ satellites deployed in Low Earth Orbit within just six years. Of these satellites, $1141$ have already been deorbited, leaving $7917$ active satellites. As such, the average operational lifespan of the deorbited satellites is approximately $2.77 $years. A comprehensive list of SpaceX’s satellites is available in the extensive work of McDowell, J.\cite{mcdowellConstellations}. The rapid expansion of the Starlink constellation has further raised concerns about the long term sustainability of LEO, prompting closer scrutiny of the quality and reliability of its publicly available orbital data. Although SpaceX regularly publishes ephemerides for each satellite, their accuracy, consistency, and value for long term analysis remain uncertain.

In this work, we evaluate the accuracy of Starlink’s publicly available data by comparing it against high fidelity orbital propagation and by training data driven models to predict satellite trajectories. This allows us to quantify the reliability of the published ephemerides, and evaluate its usability for orbit forecasting with Neural Ordinary Differential Equations, a subset of Physics-Informed Machine Learning.

\subsection{Related Work}
\vspace{0.3cm}

\noindent Over the past decade, the rise in Low Earth Orbit satellite deployments has increased research into satellite data analysis, high fidelity orbit propagation solvers, and machine learning approaches\cite{Caldas_2024}. However, in the context of Starlink, relatively few studies that analyze its public orbital data and none use it for orbit prediction.

Notably, Liu et ~\cite{LIU20243157,AIRONG2025193} analyzed the manoeuvring behavior of Starlink satellites by examining changes in the semi major axis of satellite orbits. They concluded that satellites in their operational orbits perform orbit maintenance maneuvers approximately once every one to two days. 

Their work also showed that Starlink's propagation model changes after the 48th hour. During the first two days, the ephemeris is generated using a propagator that accounts for Earth’s gravity field through spherical harmonics truncated to the 20th order and other non-conservative forces. On the third day of the propagation, however, the model is simplified, relying solely on the $J_2$ perturbation term, which captures only the effect of Earth’s oblateness.

In order to propagate a satellite’s state forward in time, numerical methods are commonly used to solve a satellite's equations of motion. Numerical integrators offer high accuracy at the cost of computational complexity. The most commonly used numerical solvers are Dormand-Prince 8(7) (RKDP8), Runge-Kutta-Nystrom 12(10) (RKN12), Adams-Bashforth-Moulton (ABM) and Gauss-Jackson (GJ) ~\cite{montenbruck2000numerical,inproceedings}. 

The equation of motion, seen in Equation \ref{eq:sat_eq_of_motion}, is separated into two coupled first order ODEs, where $\mu$ is the standard gravitational parameter, ${\mathbf  r}$ is the position vector from the satellite to Earth, $\mathbf a_{\mathrm{p}}$ represents the combined acceleration due to non-conservative forces~\cite{schaub2003analytical}.

\begin{equation}
\begin{cases}
\dfrac{d\mathbf  r}{dt} = \mathbf v, \\[1ex]
\dfrac{d\mathbf v}{dt} = -\mu \dfrac{\mathbf r}{\|\mathbf r\|^3}
  + \mathbf a_{\mathrm{p}}
\end{cases}
\label{eq:sat_eq_of_motion}
\end{equation}

Several numerical integrators have been used in this context, some specifically designed for orbital dynamics \cite{orbit_methods, aristoff2012implicit, Caldas_2024}. Such methods serve as the basis for Orekit, which is used in this study to evaluate the precision of publicly available Starlink ephemerides. Orekit is an opensource Java library for orbital mechanics and astrodynamics, widely used for high precision orbit propagation. Orekit supports detailed force models, including high order gravity harmonics, atmospheric drag, solar radiation pressure, and third-body perturbations, and has been extensively validated by independent researchers ~\cite{orekit_determination, orekit_determination_2, orekit_determination_3}. In this study, Orekit serves as the high fidelity reference for evaluating the accuracy of Starlink’s publicly available ephemerides.

To overcome limitations related to data availability and computational cost, recent studies have explored hybrid approaches that combine machine learning with physical models. Such is the work of Varey et al.~\cite{varey2024physics} that compared the performance of a traditional physics based propagator with a Neural ODE when modeling the orbit of a satellite subject to an unknown thrust profile. The Neural ODE was trained to capture deviations from the physical model by learning the thrust component directly from observational data. Alternative hybrid methods focus on correcting the errors of standard orbit propagators, instead of modeling the full trajectory itself~\cite{peng_svm,peng2019comparative,PENG2021222,PENG201944}.

\subsection{Contributions}
\vspace{0.3cm}

\noindent This work proposes a novel approach to modeling orbital parameters from real world satellite data using Neural Ordinary Differential Equations, integrating both data driven learning and physical priors to improve predictive accuracy while maintaining physical consistency. In parallel, we evaluate the quality of publicly available ephemeris data from the Starlink constellation, the largest satellite network currently in operation, by comparing its published orbital predictions against high fidelity reference trajectories generated with Orekit. Finally, the acceleration profile of conservative and non-conservative forces from Starlink's decaying satellites is extracted from the Neural Ordinary Differential Equation model.

\section{Physics-Informed Neural Networks}
\vspace{0.3cm}

\noindent Traditional Neural Networks are Machine Learning models that are able to learn complex patterns and relationships in data to make informed predictions, be it for classification or regression purposes. Each layer of a Neural Network transforms the input according to \eqref{eq:nn_layer}, where $\mathbf{W}^{(l)}$ and $\mathbf{b}^{(l)}$ are the layer's weights and biases, and $\sigma$ is a non-linear activation function.
\begin{equation}
\mathbf{h}^{(l)} = \sigma\left( \mathbf{W}^{(l)} \mathbf{h}^{(l-1)} + \mathbf{b}^{(l)} \right)
\label{eq:nn_layer}
\end{equation}

By composing these layers in succession, Neural Networks can approximate any function. Training these models involves minimizing an objective function, known as the loss function, with respect to the network's parameters using optimization algorithms. This loss function is usually the Mean Squared Error between the predicted values outputted by the Neural Network, $\hat{\mathbf{h}}(t_i)$, and the observed values, $\mathbf{h}_{\text{obs}}(t_i) $:

\begin{equation}
\mathcal{L}_{\text{data}} = \frac{1}{N} \sum_{i=1}^N \left\| \hat{\mathbf{h}}(t_i) - \mathbf{h}_{\text{obs}}(t_i) \right\|^2.
\label{eq:rmse}
\end{equation}

\noindent Optimization algorithms use \emph{backpropagation} to update the weights by computing the gradients of the loss with respect to each parameter, applying the chain rule through the composed layers. In practice, the dataset is typically divided into three disjoint subsets: a training set, a validation set, and a test set. The training set is used to fit the model parameters by minimizing the loss function, while the validation set is used to monitor generalization performance and tune parameters, such as the number of layers. Only after the optimal configuration has been found is the model evaluated on the independent test set, which provides an unbiased estimate of predictive performance. \\

\noindent In order to make accurate predictions, Neural Networks often require large amounts of data, which is often impractical in scientific applications, where data can be sparse, noisy, or expensive to obtain. Physics-Informed Neural Networks (PINNs) address this limitation by embedding physical laws directly into the training process. This prior knowledge allows the model to learn underlying dynamics with far less data, while ensuring adherence to known physical constraints.

In the context of orbital mechanics, PINNs are particularly well suited for modeling the trajectory of satellites and inferring orbital parameters from trajectory data. By enforcing the equations of motion as part of the training loss, the network can learn physical quantities, such as drag acceleration or mass.

\subsection{Neural Ordinary Differential Equations}\label{node_section}
\vspace{0.3cm}

\noindent Neural Ordinary Differential Equations (Neural ODEs or NODEs) are a subset of PINNs that operate in continuous time and continuous depth \cite{chen2018neural}. In contrast to standard PINNs, which require the explicit form of the governing differential equation, Neural Ordinary Differential Equations learn the system's dynamics directly from data. A Neural Network \(f_{\theta}\) parametrizes the time derivative of the hidden state such that:
\begin{equation}
\frac{d\mathbf{h}(t)}{dt}=f_{\theta}\bigl(\mathbf{h}(t),t\bigr) \qquad 
\mathbf{h}(t_{0})=\mathbf{h}_{0},
\label{eq:neural_ode}
\end{equation}
and an ODE solver integrates this system to obtain the trajectory \(\hat{\mathbf{h}}(t)\). These outputs are then compared to the observed data, $\mathbf{h}_{\text{obs}}(t)$, using the loss function \eqref{eq:rmse}.
\smallskip

\noindent Backpropagating through every solver step would make the memory cost increase at every step. Instead, Neural ODEs rely on the \emph{adjoint sensitivity method}. This method uses an adjoint system of ODEs that is integrated backward in time to obtain the loss sensitivities:
\[
\frac{\partial \mathcal{L}}{\partial t_0}, \qquad
\frac{\partial \mathcal{L}}{\partial t_1}, \qquad
\frac{\partial \mathcal{L}}{\partial \mathbf{h}(t_0)}, \qquad
\frac{\partial \mathcal{L}}{\partial \theta}.
\]

\noindent This system is defined with the adjoint variable:

\begin{equation}
    \mathbf{a}(t) = \frac{\partial \mathcal{L}}{\partial \mathbf{h}(t)},
\end{equation}

\noindent which is itself governed by an ODE, that describes how sensitive the loss is with respect to the state:
\begin{equation}
\frac{\partial\mathbf{a}(t)}{\partial t}
  = -\mathbf{a}(t)\frac{\partial f_\theta(\mathbf{h}(t),t)}{\partial\mathbf{h}}.
\label{eq:adjoint_ode}
\end{equation}

\noindent The necessary ODEs are computed together as the adjoint system, $ad(t)$, with a single call to an ODE solver:

\begin{equation}
\frac{d\,\mathrm{ad}(t)}{dt} = -
\begin{bmatrix}
\mathbf{a}(t)^{\!\top}\,\dfrac{\partial f_\theta}{\partial \mathbf{h}}\, \\[6pt]
\mathbf{a}(t)^{\!\top}\,\dfrac{\partial f_\theta}{\partial \theta}\ \\[6pt]
\mathbf{a}(t)^{\!\top}\,\dfrac{\partial f_\theta}{\partial t}\ 
\end{bmatrix}(\mathbf{h}(t),t),
\label{eq:adjoint_aug}
\end{equation}

\noindent where the solutions are the loss gradients.
This method returns the exact gradients while requiring memory that scales with the number of parameters, not the number of solver evaluations, making it more suitable than backpropagation.

In this study, we use a Neural ODE model with the adjoint sensitivity method to learn the orbital dynamics of satellites from trajectory data, with the goal of estimating latent physical quantities.

\subsection{Augmented Neural Ordinary Differential Equations}
\vspace{0.3cm}

\label{subsec:anodes}

\noindent
Plain Neural ODEs learn a continuous time flow
\(\phi_t : \mathbb{R}^{d}\!\to\!\mathbb{R}^{d}\) by solving an initial value
problem.  The \emph{Picard--Lindelöf} theorem guarantees a unique local solution
whenever the vector field \(f\) is continuous in \(t\) and follows the \emph{Lipschitz} condition, i.e.\ there exists a constant \(L\ge 0\)
such that
\begin{equation}
\label{eq:lipschitz}
\|f(t,\mathbf h_1)-f(t,\mathbf h_2)\|
\;\le\;
L\,\|\mathbf h_1-\mathbf h_2\|,
\end{equation}

\noindent
for all \(t\) in some interval \(I\) and all
\(\mathbf h_1 \text{ and }\mathbf h_2\) in a region \(U\subset\mathbb R^{d}\)\cite{teschl2012ordinary}.\\

\noindent This means that two distinct initial conditions can never meet at the same state at the same time. This restriction limits plain NODEs from representing mappings that require trajectory intersections or folding, such as when non-conservative forces cause different initial states to converge. However, the constraint can be lifted by using Augmented NODEs \cite{dupont2019augmentedneuralodes}. Augmented Neural ODEs lift the state
\(\mathbf h\in\mathbb R^{6}\) to an augmented state
\(\mathbf h^*=[\mathbf h;\,\mathbf a]\in\mathbb R^{6+k}\) with auxiliary
coordinates \(\mathbf a(0)=\mathbf 0\).
The unique flow in the augmented space is projected back to
\(\mathbb R^{6}\), allowing the intersection while retaining well formed ODE solutions.

%\subsection{Orbit Prediction}

%To prevent potential collisions between satellites, their trajectories need to be accurately %predicted in the near future. Several methods for orbit prediction have been developed, %which generally fall into one of three categories: analytical methods, numerical methods, %and data-driven methods.

%Numerical methods focus on solving the satellite's equations of motion with numerical %integrators. These methods are widely used in orbit prediction due to their high accuracy. %However, they can be computationally intensive, particularly for long-term forecasts. %Specialised numerical integration solvers have been developed specifically for orbit %prediction tasks %\cite{montenbruck1992numerical}\cite{jones2012survey}\cite{berry2004gaussjackson}\cite{aristoff2012implicit}.

\section{Data source}
\vspace{0.3cm}

\noindent Since $2019$, SpaceX has provided downloadable ephemeris files for each Starlink satellite on the Space-Track website. These files include tabular data with the predicted positions and velocities over time, along with associated covariance matrices and trajectory metadata.

Each ephemeris file forecasts a three day trajectory using a combination of propagation models, the specifics of which have not been disclosed by SpaceX. The positions and velocities are provided in kilometers and kilometers per second, respectively, and are expressed in the MEME (Mean Equator Mean Equinox) J2000.0 reference frame. The covariance values are given in the UWV frame, which is equivalent to the RTN (Radial, Transverse, Normal) frame. The format specifications for ephemeris files are outlined in the Space-Track Operator Handbook~\cite{spacetrackHandbook}. This handbook includes all the file naming specifications, data field descriptions, reference frames, and units.

The ephemeris data is refreshed every eight hours, with previous versions deleted after 24 hours. This limited retention window, combined with the lack of transparency regarding the propagation method, has reduced the utility of the data for research purposes. Despite this, the dataset remains one of the only publicly available sources of orbital predictions for satellites and offers an interesting opportunity to better understand Starlink's orbital dynamics and operational behaviour. 

Given the limited time window to obtain each ephemeris file, the data was collected using Space-Track's API every eight hours. The collection of data started on the 28th of November $2024$ and ended on the 8th of January $2025$. Due to the amount of data, only $1500$ satellites were collected.

\section{High Fidelity Propagator}
\vspace{0.3cm}

\noindent Given the lack of transparency surrounding the propagation models used in Starlink’s ephemeris files, it was first necessary to assess the quality of the data before using it as input to a Neural Network. To this end, the ephemerides were compared against a high-fidelity orbital propagator. In this study, Orekit was used to propagate Starlink satellites based on initial state vectors. The resulting trajectories served as a reference to evaluate the accuracy and consistency of Starlink's published ephemerides.

\subsection{Satellite Classification}

\noindent
To ensure a fair evaluation of Starlink’s propagation accuracy, only satellites classified as \textit{Stable} were included in the Orekit comparison. This filtering step was essential to minimize the influence of orbital decay, which introduces additional variability unrelated to the fidelity of the propagation model.

Starlink satellites were grouped into three categories based on their orbital behaviour: \textit{Stable}, \textit{Deorbiting}, and \textit{Decayed}.

\begin{figure}[H]
\centering
\includegraphics[width=\columnwidth]{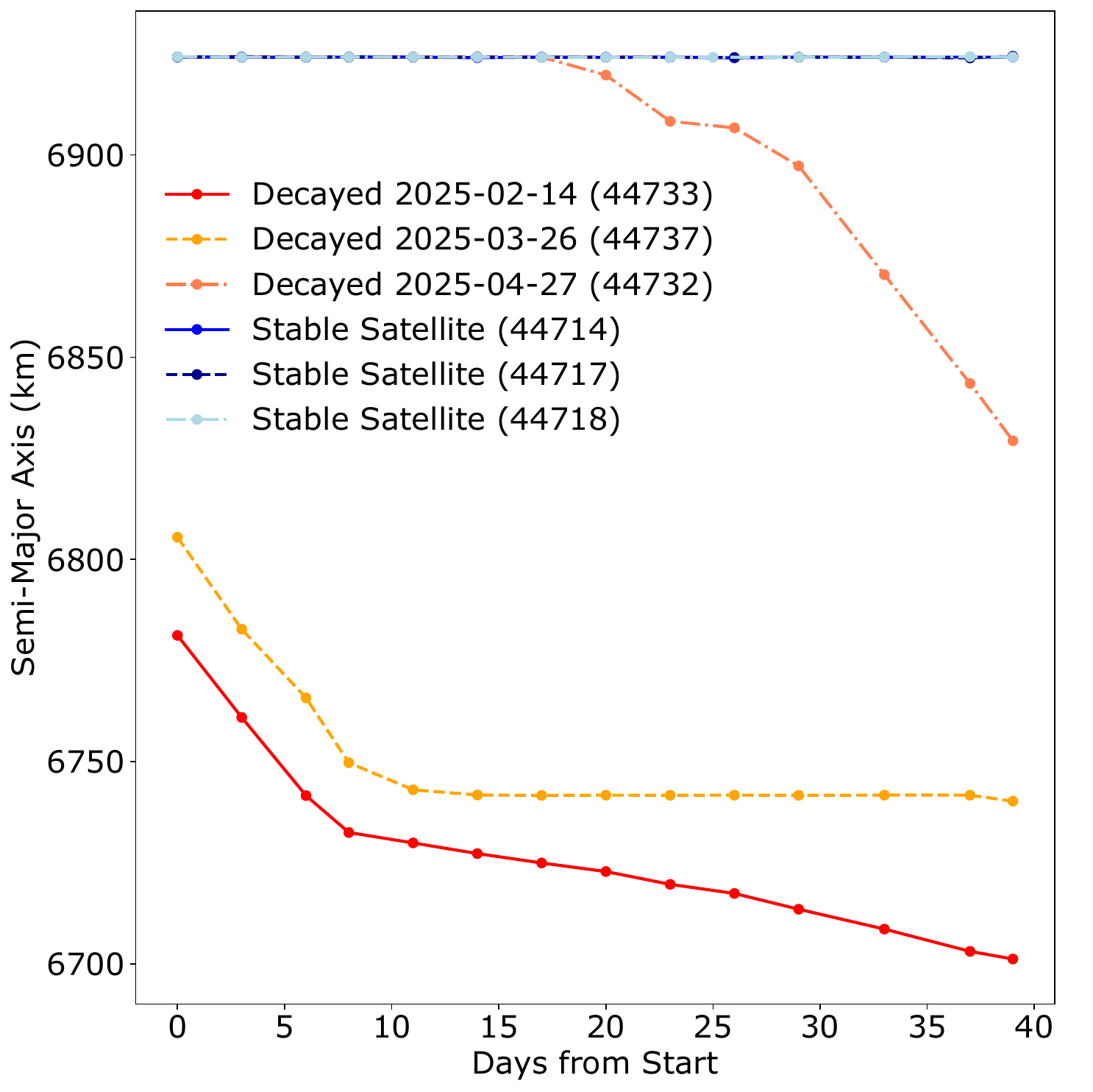}
\caption{Graph showing the evolution of satellites' semi major axis throughout the data collection period. The semi major axis values were recorded for each orbit every $2$ days and represent the mean value during that time. The blue toned lines correspond to satellites that remained relatively stable, with minor variations due to periodic manoeuvres. The orange-toned lines represent satellites that began deorbiting and ultimately reentered the atmosphere after the collection period.}
\label{fig:sma_evol}
\end{figure}

Satellites were labeled as \textit{Deorbiting} if their semi major axis exhibited a consistent downward trend during the observation period and if they decayed shortly after the data collection window ended. This criterion helped distinguish genuine decay trajectories from temporary altitude adjustments due to manoeuvres. Figure~\ref{fig:sma_evol} shows an example of such deorbiting satellites, highlighted in orange tones.

The \textit{Decayed} category includes satellites that fully re-entered the atmosphere during the study period, as reported by Space-Track. Out of the $1,500$ satellites tracked, $58$ decayed over the two-month collection window, corresponding to an average decay rate of approximately $1.26$ satellites per day.

Satellites whose semi major axis remained approximately constant throughout the observation window were classified as \textit{Stable}. These are shown in blue tones in Figure~\ref{fig:sma_evol}. Although considered stable, these satellites still performed periodic orbit maintenance manoeuvres — typically once every one or two days~\cite{LIU20243157,AIRONG2025193}. Additionally, small oscillations in their semi major axis were observed due to non-conservative perturbations such as atmospheric drag.

\subsection{Parameter discovery}
\vspace{0.3cm}

\noindent To initialize Orekit’s orbit propagation, a set of satellite specific parameters is required. These include the initial state vectors (position and velocity), their uncertainties, the satellite’s mass, the propagation start time, and a set of perturbative forces along with their associated coefficients. However, several of these, particularly the physical parameters governing drag, reflectivity, and third-body interactions, are not publicly disclosed by Starlink and can vary depending on orbital regime.

To allow for a fair comparison with the published ephemerides, these unknown parameters were estimated via Bayesian optimization. The objective was to minimize the Root Mean Square Error between Orekit’s propagated trajectory and Starlink’s predicted state vectors. Due to the stochastic nature of the optimizer, each satellite underwent $50$ independent optimization runs, and the parameter set yielding the lowest RMSE was retained.

The force models considered during this discovery phase included:
\begin{itemize}
    \item Atmospheric drag, with the drag coefficient $C_D$ treated as a free parameter;
    \item Solar radiation pressure, with a free reflectivity coefficient $C_R$;
    \item Third-body perturbations from the Moon and Sun, toggled on/off;
    \item Satellite radius, under the assumption of a spherical geometry.
\end{itemize}

To improve robustness, a single three day trajectory was constructed for each satellite by stitching together nine consecutive ephemeris files. Since the ephemerides are updated every eight hours, each file contains an overlapping 64 hour window with its predecessor. By stitching together the first 8 of these higher confidence segments, a more reliable three day trajectory can be constructed with reduced numerical error accumulation. The starting date for each trajectory was randomly sampled from within the data collection period to ensure diversity.

\subsection{High Fidelity Comparision}
\vspace{0.3cm}

\begin{figure}[H]
\centering
\includegraphics[width=\columnwidth]{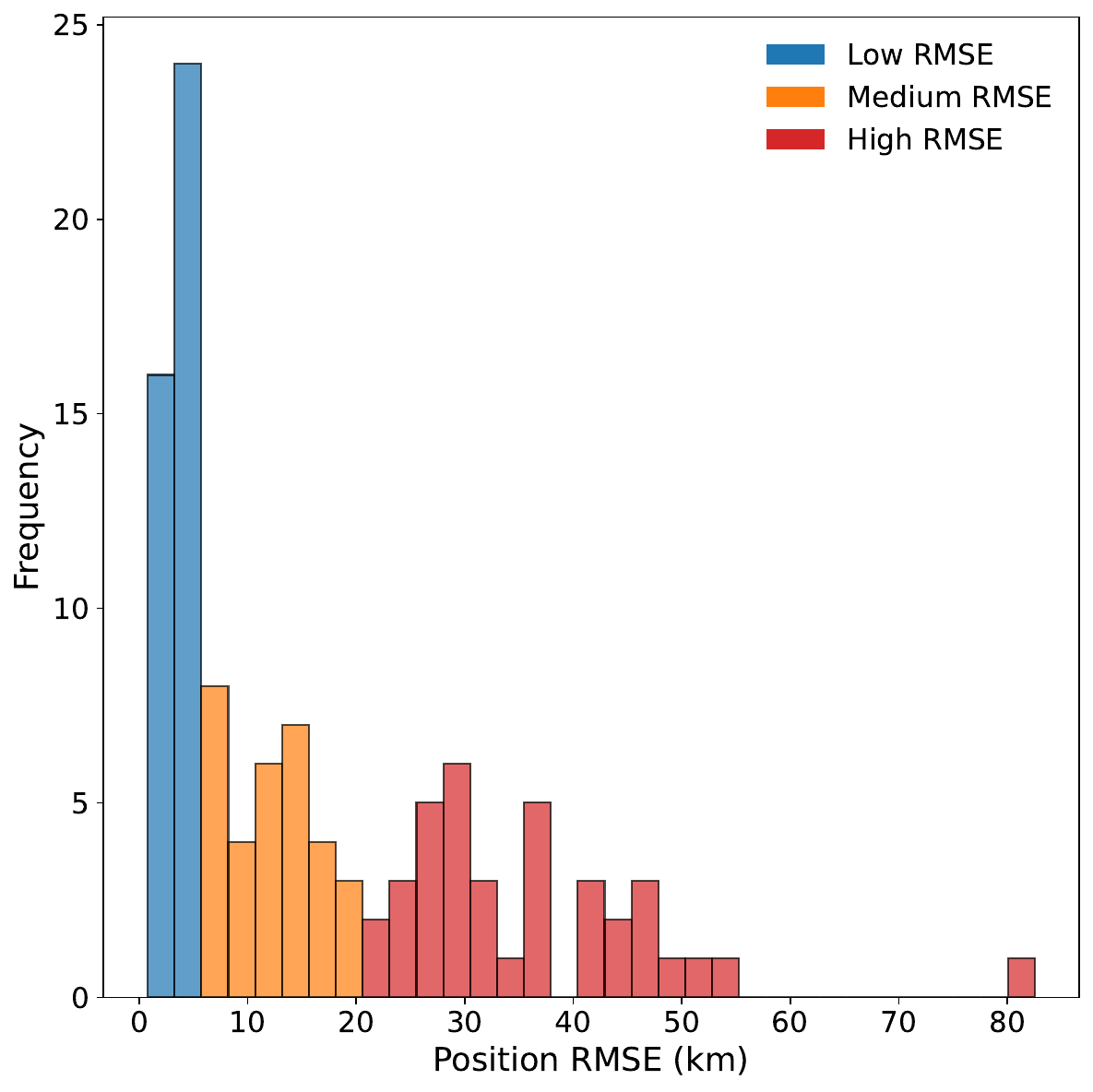}
\caption{Histogram of the RMSE values between the Orekit Propagation and the Starlink ephemerides of $300$ satellites. These values were separated into three equal sized groups based on their RMSE.}
\label{tab:distribution_rmse}
\end{figure}

\noindent Using the estimated parameters for each of $300$ satellites, positions obtained from Orekit’s high fidelity propagator were compared against the published Starlink ephemerides. The resulting RMSE values were classified into three categories of equal size, as seen in Table~\ref{tab:rmse_compact}. While many satellites exhibit relatively small deviations, others show larger discrepancies. As illustrated in Figure~\ref{tab:distribution_rmse}, the error distribution is right skewed, with a pronounced long tail extending toward higher RMSE values.

\begin{table}[H]
\centering
\caption{Satellite RMSE Classification (Position) across 300 satellites.}
\begin{tabular}{l c c}
\hline
\textbf{Category}  & \textbf{Error Range [m]} & \textbf{Mean Error[m]} \\
\hline
Low    &  742.31 -- 5,277.68  & 3,098.64 \\
Medium &  5,281.48 -- 11,361.56 & 7,828.00 \\
High   &  11,586.48 -- 82,510.46 & 27,812.51 \\
\hline
\end{tabular}
\label{tab:rmse_compact}
\end{table}

\noindent A component wise analysis showed that the $y$ coordinate was the most error prone, accounting for the dominant error in 155 satellites (51.7\%), compared to $130$ $(43.3\%)$ for $x$ and only $15$ $(5.0\%)$ for $z$, as seen in Table~\ref{tab:component_rmse}.

\begin{table}[H]
\centering
\caption{Position error component analysis across $300$ satellites. Includes the frequency of when each component has the highest error, in order to understand which component of the position is harder to model.}
\begin{tabular}{lcc}
\hline
\textbf{Component} & \textbf{Frequency} & \textbf{Mean Error [m]} \\
\hline
X & 130 (43.3\%) & 7396.35 \\
Y & 155 (51.7\%) & 7424.86 \\
Z & 15  (5.0\%)  & 7323.98 \\
\hline
\end{tabular}
\label{tab:component_rmse}
\end{table}

\begin{figure}[H]
\centering
\includegraphics[width=1.0\columnwidth]{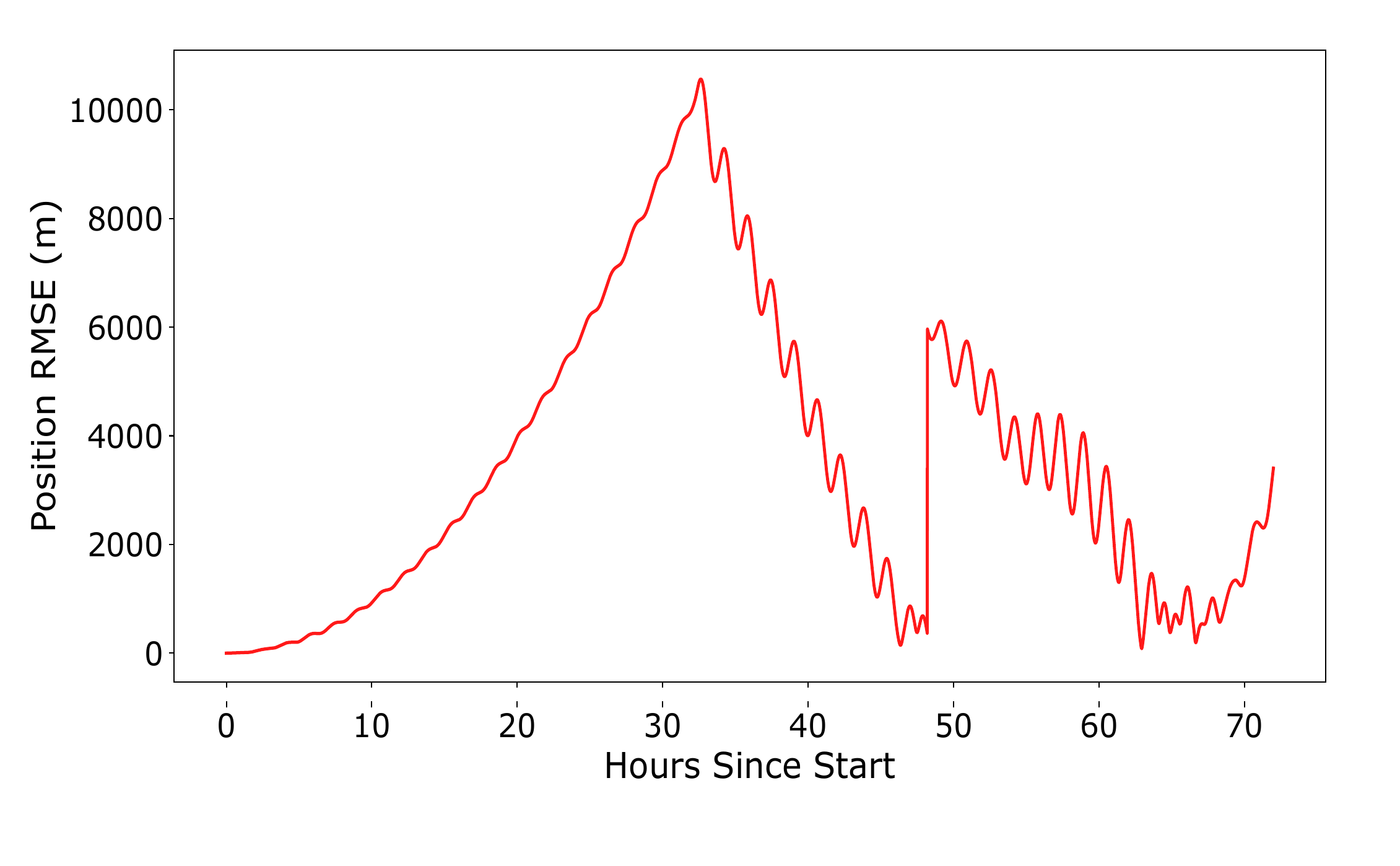}
\caption{Evolution of position RMSE for Satellite $48104$. The peak in the plot marks a $\sim$1.9 h manoeuvre, during which both RMSE and covariance determinants spike sharply. A second marked change occurs near the $48$ h mark, consistent with a switch in Starlink’s internal propagation model.}
\label{fig:rmse_48104_orekit}
\end{figure}

\begin{figure}[H]
\centering
\includegraphics[width=1.0\columnwidth]{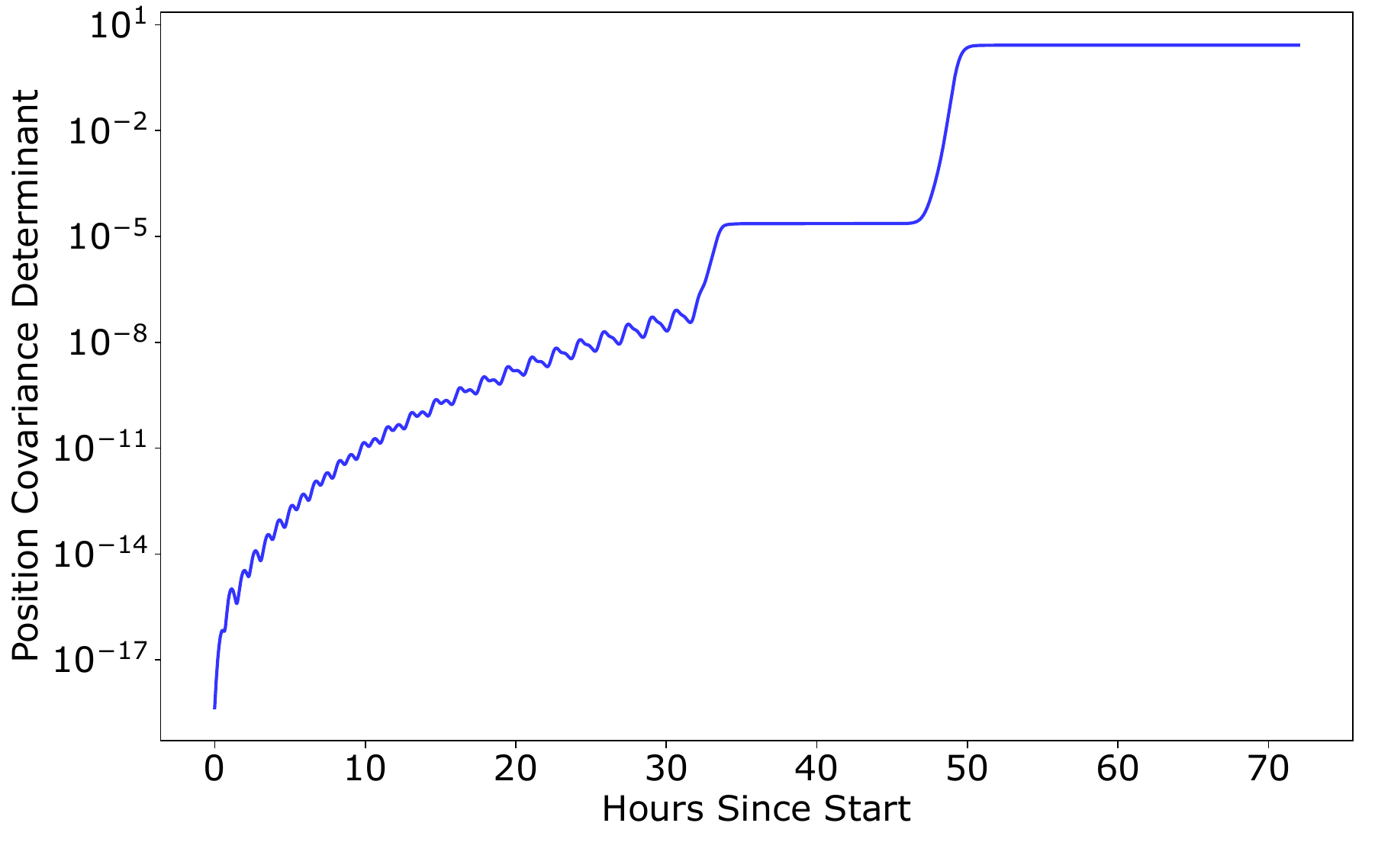}
\caption{Evolution of position covariance determinant for Satellite $48104$. The peak in the plot marks a $\sim$1.9 h manoeuvre, during which both RMSE and covariance determinants spike sharply. A second marked change occurs near the $48$ h mark, consistent with a switch in Starlink’s internal propagation model.}
\label{fig:cov_48104_orekit}
\end{figure}

Several factors contribute to the observed high RMSEs between Starlink and Orekit trajectories:

\begin{enumerate}
  \item \textbf{Unmodelled manoeuvres.} Starlink satellites perform frequent orbit maintenance manoeuvres, which are not accounted for in the Orekit propagator. These result in sudden trajectory changes that Orekit cannot replicate, leading to sharp spikes in RMSE at the time of the manoeuvre (Figures~\ref{fig:cov_48104_orekit} and~\ref{fig:rmse_48104_orekit}).
  
  \item \textbf{Simplified internal propagation models.} Starlink appears to switch its internal propagator at the 48-hour mark. Beyond it, only the gravitational force is retained. This switch introduces a discontinuity in the trajectory causing divergence from Orekit's high fidelity propagator.

  \item \textbf{Intrinsic ephemeris error.} Starlink’s own predictions accumulate errors over time. This can be estimated by comparing overlapping segments of successive ephemeris files. This internal error grows with forecast horizon and varies across orbital regimes.
\end{enumerate}

\noindent
An example of manoeuvre induced divergence is shown in Figures~\ref{fig:cov_48104_orekit} and~\ref{fig:rmse_48104_orekit}, where both position RMSE and the determinant of the covariance matrix spike simultaneously during a $\sim$2-hour manoeuvre window. A second spike at 48~h coincides with the suspected switch in Starlink’s internal propagation model.

\noindent
To evaluate Starlink's intrinsic prediction error, each three day ephemeris was compared to overlapping segments from newer forecasts. This approach assumes that the first 8~hours of each file are the most accurate, and that subsequent files incorporate more recent and accurate orbital information. The error between overlapping forecasts thus provides a proxy for Starlink's own propagation uncertainty. Most satellites show a distinct jump in residuals near 48~h, whereas decaying satellites exhibit a smoother increase, suggesting the use of different internal models for different orbital phases.

\noindent
To further evaluate the consistency of the errors observed, each satellite's RMSE category from the Orekit comparison was compared to its RMSE category based on Starlink’s own internal overlap error. A high degree of agreement was found between the two, suggesting that some of the discrepancies in Orekit’s results stem from limitations in the Starlink ephemerides themselves.

\noindent
\begin{itemize}
    \item Of the $100$ satellites classified as \emph{Low error} in the Orekit-based RMSE analysis, $81$ were also classified as \emph{Low error} using Starlink's internal overlap metric.
    \item In the \emph{Medium error} group, $95$ of $100$ satellites were consistently classified under both methods.
    \item In the \emph{High error} group, $72$ of $100$ satellites matched.
\end{itemize}

\noindent
This strong alignment reinforces the conclusion that a significant portion of the RMSE observed in the Orekit comparison arises from inconsistencies and limitations in the publicly released Starlink ephemerides.

\noindent
It is important to note that the Orekit analysis is not intended to serve as a performance baseline for subsequent machine learning models. Instead, it highlights the inherent variability and limitations in the source data. When evaluating the Neural ODE results, lower than expected performance may reflect the quality of the training data as much as the model itself.

\section{Neural Ordinary Differential Equations for Modeling Orbital Dynamics  }\label{sec:nodemodel}
\vspace{0.3cm}

\noindent This section presents the formulation of the Neural  Ordinary Differential Equation model used to propagate Starlink satellites, with particular emphasis on deorbiting satellites in order to extract their acceleration profiles.\\

\noindent In orbital mechanics, a satellite’s state vector 
\[
\mathbf h(t) = \begin{bmatrix}\mathbf r(t) \\[4pt] \mathbf v(t)\end{bmatrix} \in \mathbb{R}^6,
\]  
comprising of position $\mathbf r(t)$ and velocity $\mathbf v(t)$ vectors, evolves according to the second order ordinary differential equation shown in \eqref{eq:sat_eq_of_motion}. Neural ODEs applied to orbital mechanics approximate these ODEs directly from observed trajectory data, learning a continuous time model of the dynamics, as discussed in Section \ref{node_section}.  \\

In order to facilitate the learning of the physics involved in orbital mechanics, we split the dynamics into a \emph{known} physics term and a \emph{learnable} perturbation term. This way, the model only needs to learn part of acceleration profile, $\mathbf a_{\mathrm{pert},\theta}(\mathbf h, t)$. The derivative of the position does not need to be learned by the Neural Network, as it can simply be set to be equal to the velocity. The learned perturbation is then incorporated into the gravitational acceleration alongside the $J_2$ and $J_3$ terms. Together with the Newtonian gravity, these contributions define the base acceleration,
\begin{equation}
\mathbf a_{base} =
-\mu\frac{\mathbf r}{\|\mathbf r\|^3}
+\mathbf a_{J_2} + \mathbf a_{J_3},
\label{eq:ode}
\end{equation}

\noindent where $J_2$ and $J_3$ denote the second and third zonal harmonics of Earth’s gravitational potential. \\

\begin{figure}[H]
\centering
\includegraphics[width=0.8\columnwidth]{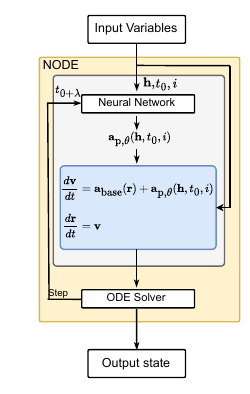}
\caption{Schematic of the Neural ODE architecture used for satellite trajectory prediction. The input variables consist of the satellite's initial state $\mathbf h$ and the prediction time, along with the initial angular momentum, space weather features and short history of past states. The Neural Network outputs the acceleration, $a_p$, which is then added to the two body acceleration with $J_2$ and $J_3$ correction ($a_{total}$). This output is integrated over time using an ODE solver to produce the predicted state at a future time step. The ODE solver queries the Neural Network at each integration step ($\lambda$) to obtain the time derivatives, which are then used to update the state vector. The process is repeated until the prediction time is reached. The output of the NODE is the predicted position and velocity of the satellite at the specified time step.}
\label{fig:node_architecture}
\end{figure}

In the framework used in this paper, the total state derivative comprising of the velocity (the derivative of position) and the sum of the gravitational and learned perturbative accelerations (the derivative of velocity), are passed to a numerical ODE solver along with the initial state. Whenever the integrator needs to take a step $(\lambda)$, it asks the Neural Network $a_{\mathrm{pert},\theta}(\mathbf h, t,i)$ for the derivatives. At each evaluation, this Neural Network receives as input the satellite’s current position and velocity and the prediction time, along with a short history of past positions, velocities, and space weather parameters, denoted as $i$. After taking the necessary steps, the solver returns the predicted state for the requested time. A diagram of the NODE architecture can be seen in Figure~\ref{fig:node_architecture}.

The input values to the Neural Network have different scales. The position is left in km, meaning it's values are in the order of $10^3$ and the velocity is left in km/s with values in the order fo $10^1$. Feeding these raw values straight into the network makes it harder to solve the optimization problem. Because weight updates in gradient based methods are proportional to the input magnitude, large scale features dominate. The position features will push its associated weights to grow or shrink far more aggressively than the smaller scale features. This means that when optimizing the Neural Network to output the most correct results, more steps are wasted to correct that imbalance instead of discovering useful patterns. In order to avoid this, normalizing the input features is necessary. In order to ensure all features lied in the same range, they were all rescaled to be in the interval $[-1,1]$. The normalization was performed as follows:

\begin{align}
\mathbf u'  &= 2 \cdot \frac{\mathbf u_{\text{dataset}} - \mathbf u_{\min}}{\mathbf u_{\max} - \mathbf u_{\min}} - 1 ,\\
\label{normalize}
\end{align}

\noindent where $\mathbf u'$ is the normalized variable used in the model. The minimum and maximum values used for normalization correspond to the minimum and maximum values of each feature in the dataset used to train the model.
The normalization of $t$ was done differently. Instead, it was rescaled to be in the interval $[0,2\pi]$. This was done so that $t$ becomes the target angular position $\theta$, representing the point in the orbit for which the prediction is desired. If the requested state corresponds to an angle beyond the training domain $(e.g., \theta=3\pi)$, the model can be called iteratively.

\subsection{Model Training}
\vspace{0.3cm}

\noindent To train the Neural ODE, the Starlink ephemerides were divided into disjoint training, validation, and test sets. Two models were trained, one with only orbits from decaying satellites and another with only orbits from stable satellites. The second model was only used to compare how the model performs under different orbital phases.

\paragraph{Model specifications}
The NODE architecture comprised a Neural Network with $3$ hidden layers, each with $256$ neurons and Tanh activations. The Tanh (hyperbolic tangent) activation function maps inputs to the range $[-1,1]$, providing smooth gradients and helping to stabilize training by avoiding unbounded growth in the hidden representations.
Together with a dropout rate of $0.05$, training was performed for $1000$ epochs with the Adam optimizer. Each training orbit contributed $96$ uniformly sampled points. Numerical integration of the ODE was performed with the 4th-order Runge–Kutta method ~\cite{hairer1993solving}. Additionally, $6$ augmented dimensions were added to input.\\

The models were trained with $60$ orbits, validated on $30$ and tested on $30$. Training of the Neural Network weights is driven by the mean squared error loss, as mentioned in Section~\ref{node_section}, between the predicted states and the observed ephemeris. We split this into a position loss and a velocity loss, then weight the position term $1.8$ times as heavy to compensate for the network’s greater difficulty in learning accurate positions. The network depth, the number of orbits and samples per orbit, and all other training parameters were selected by optimizing for the lowest Root Mean Squared Error on the validation orbits. Evaluation was carried out on the test set.\\

\begin{figure}[H]
\centering
\includegraphics[width=1\columnwidth]{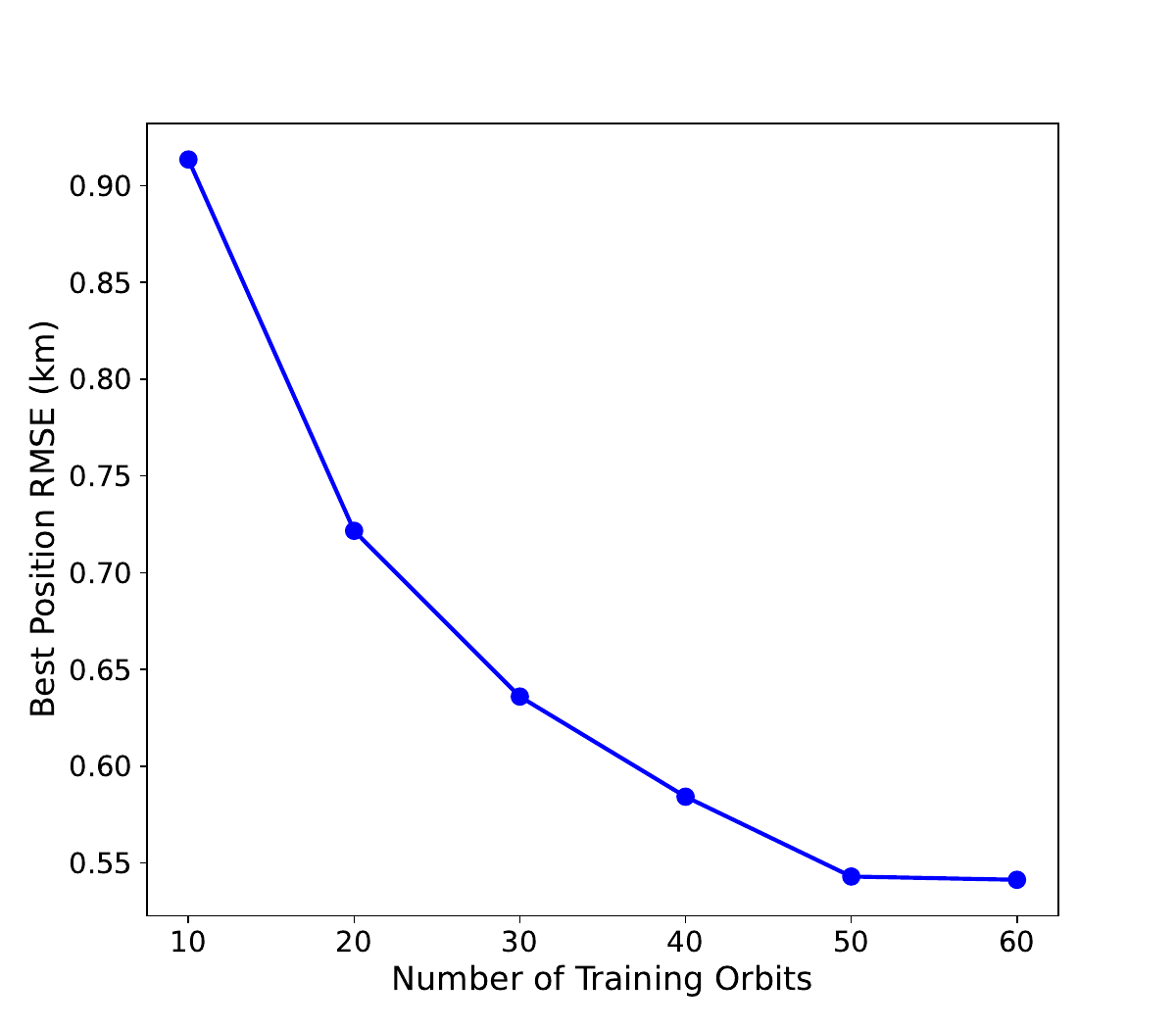}
\caption{Figure showing the evolution of the models position RMSE for validation orbits. The error decreases steadily from $0.91$~km with $10$ orbits to $0.54$~km with $60$ orbits.}
\label{fig:orbits_rmse}
\end{figure}

The number of training orbits changed the performance of the models significantly. 
Figure~\ref{fig:orbits_rmse} illustrates how the model trained on deorbiting satellites improved its accuracy on the validation set as the number of training orbits increased. With only $10$ orbits, the mean position RMSE was $0.91$~km, while training with $60$ orbits reduced the error to $0.54$~km. Beyond $50$ orbits, however, the 
improvement was marginal (approximately $30$~m). Training time scaled linearly with dataset size, 
rising from about $2$~hours for the $10$ orbit model to $16$~hours for the $60$ orbit model.

\begin{figure}[H]
\centering
\includegraphics[width=1\columnwidth]{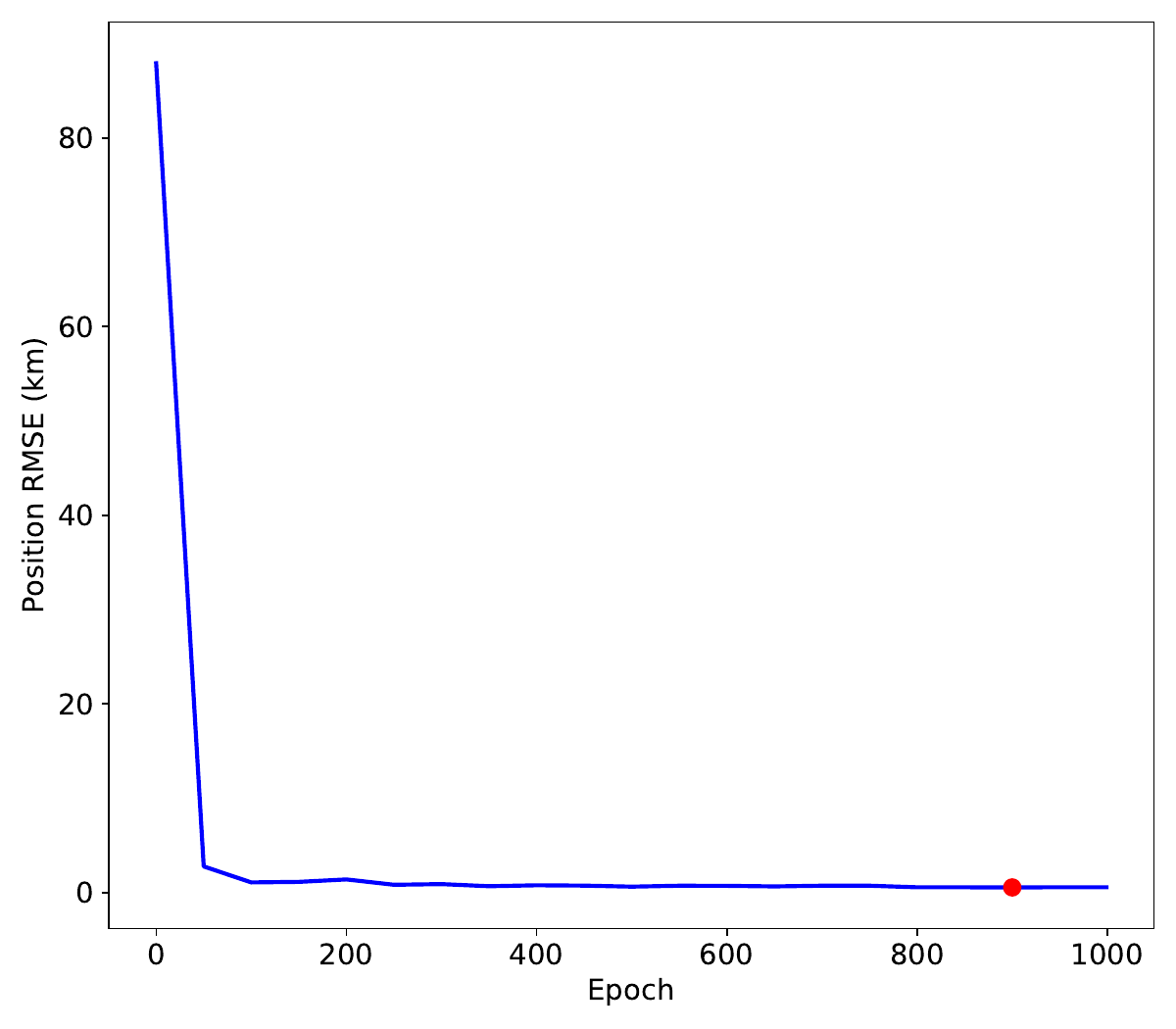}
\caption{Evolution of the position RMSE during training of the Neural ODE. 
The initial error was $87.9$~km, which decreased to $0.56$~km by the final epoch. 
The best performance was achieved at epoch~$900$, with a position RMSE of $0.54$~km.}
\label{fig:position_rmse_evolution}
\end{figure}

Figure~\ref{fig:position_rmse_evolution} shows the evolution of the position RMSE 
over the course of training. The deorbiting model was trained for $1000$ epochs, with the validation error decreasing during the early stages and then gradually converging. The lowest position RMSE was achieved at epoch~$900$ ($0.54$~km), after which performance remained stable without further improvement. By the final epoch, the error had slightly increased to 0.56~km, indicating that the model began to plateau and that longer training did not yield significant gains.

\section{Results and Discussion}
\vspace{0.3cm}

\noindent
This section presents the outcomes of the model described in Section~\ref{sec:nodemodel} when applied to deorbiting satellites. Additionally, the learned non-conservative acceleration profile is examined and compared against stable satellites.\\

The Neural ODE was trained and evaluated on satellites classified as \emph{Deorbiting}, since these provide the strongest signatures of non-conservative effects during orbital decay. Another model was trained on \emph{Stable} satellites for comparison, in order to benchmark the learned acceleration profile against cases with minimal decay. Each trajectory corresponded to a single orbit, sampled at 96 points.\\

\begin{figure}[H]
\centering
\includegraphics[width=\columnwidth]{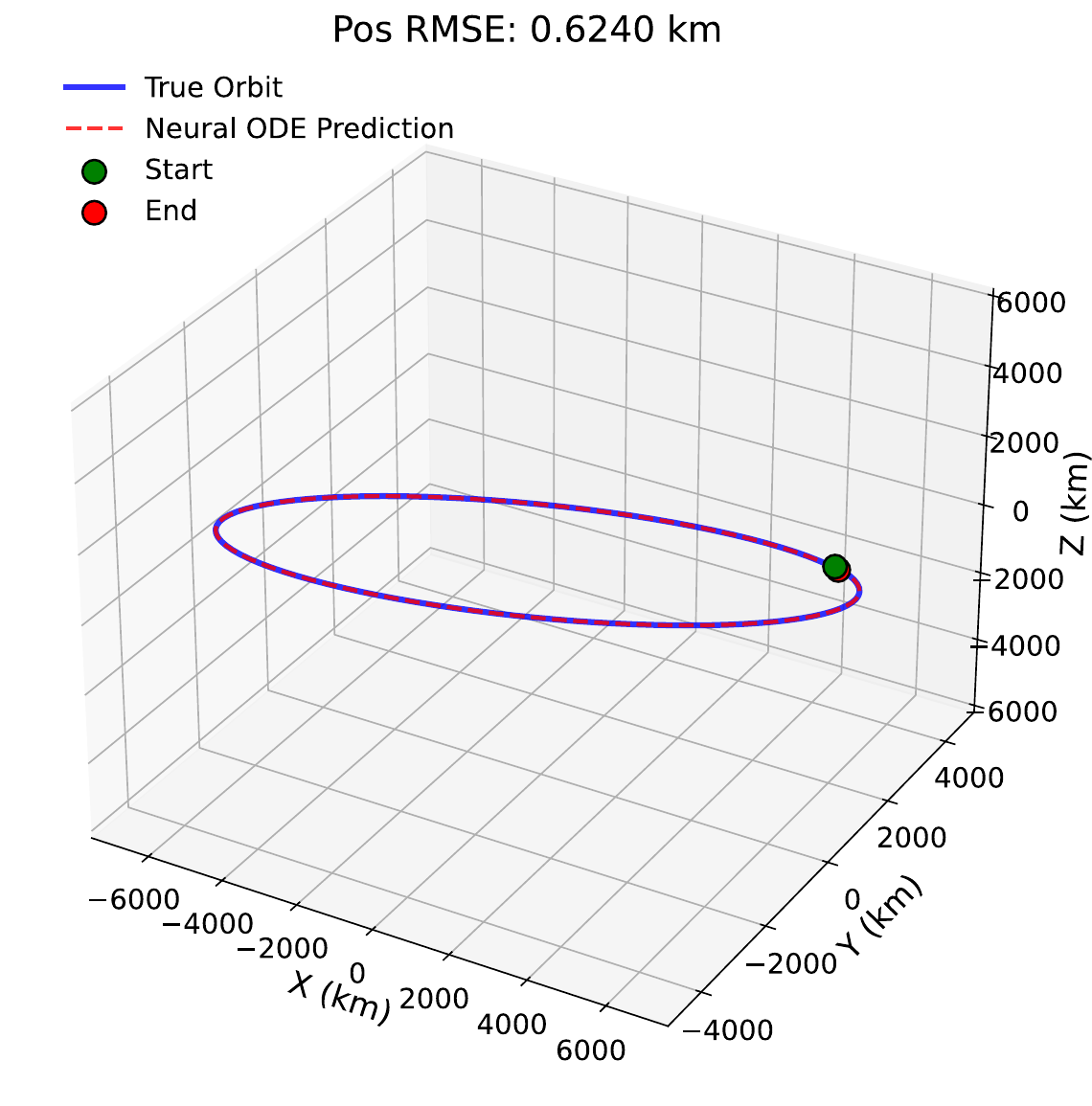}
\caption{Neural ODE prediction of one orbit for a \emph{Deorbiting} satellite (ID~$47163$), which re-entered the atmosphere on 2025-01-23. The model reconstructs the trajectory using only the learned perturbation acceleration component combined with the base gravitational terms. The resulting position RMSE for this satellite is $0.624$~km, which is consistent with the average error observed across the deorbiting dataset.}

\label{fig:3_d_orbit_sat_decay}
\end{figure}

For stable satellites, the Neural ODE achieves a mean position RMSE of
approximately $0.30$~km, with a minimum error of $0.06$~km and a maximum of
$0.88$~km. Velocity errors are correspondingly small, averaging
$3.05\times10^{-4}$~km/s, as seen in Table~\ref{tab:rmse_combined}. In contrast, deorbiting satellites exhibit larger discrepancies, as expected given the stronger perturbations and greater trajectory variability during decay. The mean position RMSE rises to $0.60$~km, with values between $0.10$~km and $1.35$~km, while the mean velocity RMSE approximately doubles to $6.10\times10^{-4}$~km/s. An example of an orbit from a deorbiting satellite modeled by the Neural ODE is shown in Figure~\ref{fig:3_d_orbit_sat_decay}, with a position RMSE of  $0.624$~km, close to the average for this category.

\begin{table}[H]
\centering
\caption{RMSE of the Neural ODE over one orbit for stable and deorbiting satellites.}
\begin{tabular}{lccc}
\hline
\textbf{Category} & \textbf{Metric} & \textbf{Mean} & \textbf{Std} \\
\hline
Stable & Position [km]   & 0.302 & 0.242 \\
       & Velocity [km/s] & $3.05\times10^{-4}$ & $2.47\times10^{-4}$ \\
\hline
Deorbiting & Position [km]   & 0.600 & 0.408 \\
            & Velocity [km/s] & $6.10\times10^{-4}$ & $4.30\times10^{-4}$ \\
\hline
\end{tabular}
\label{tab:rmse_combined}
\end{table}

\begin{figure}[H]
\centering
\includegraphics[width=1\columnwidth]{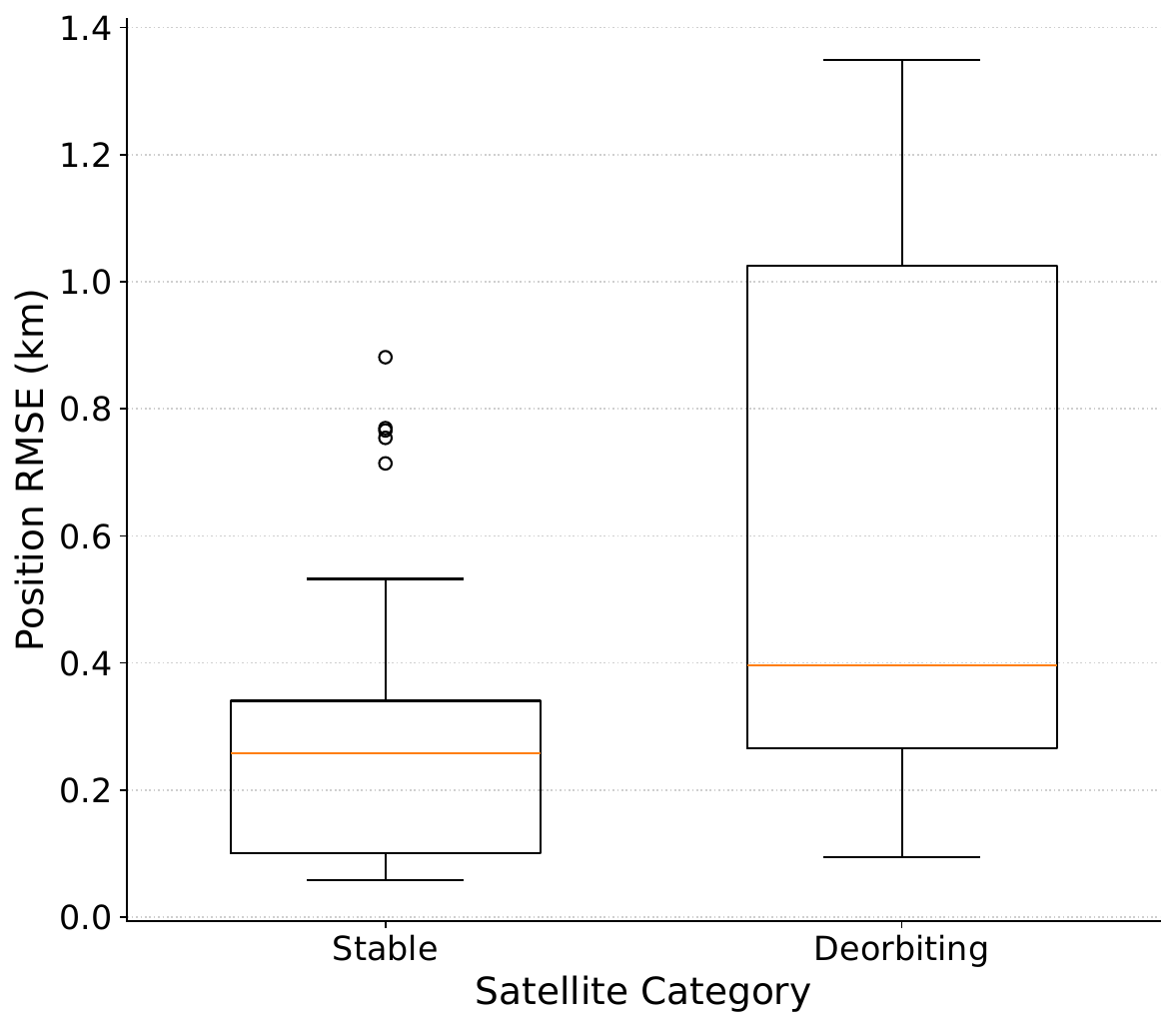}
\caption{Per orbit position RMSE for Stable vs. Deorbiting satellites. The deorbiting group shows a higher median ($\sim$0.40\,km vs.\ $\sim$0.23\,km) and more variable errors.}

\label{fig:position_rmse_boxplot}
\end{figure}

Figure~\ref{fig:position_rmse_boxplot} compares the per orbit position RMSE of the Neural ODE for \emph{Stable} and \emph{Deorbiting} satellites. Deorbiting cases show a higher median error and greater variability, with a wider spread. In contrast, stable satellites exhibit tighter error bounds with only a few outliers. These results are consistent with Table~\ref{tab:rmse_combined}, confirming that trajectory prediction is more challenging during orbital decay.

Following the evaluation of the model’s performance in predicting satellite trajectories, the learned acceleration profile can be extracted. As illustrated in Figure~\ref{fig:node_architecture}, the model is trained to output a non-conservative component of the perturbation acceleration, $a_{\mathrm{pert},\theta}$.
\\

For deorbiting satellites, the mean acceleration magnitude of this learned perturbation is $2.97\times10^{-8}\,\mathrm{km/s^2}$, nearly twice that of stable satellites ($1.66\times10^{-8}\,\mathrm{km/s^2}$). This increase reflects the stronger perturbations of orbital decay. A component wise breakdown further reveals different dominant directions: in deorbiting satellites, the $z$ component contributes the largest share $(42\%)$, whereas for stable satellites the acceleration is predominantly aligned with the $x$ component $(53\%)$. These results show that the model learns distinguishable non-conservative profiles for satellites in different orbital regimes.

\begin{figure}[H]
\centering
\includegraphics[width=\columnwidth]{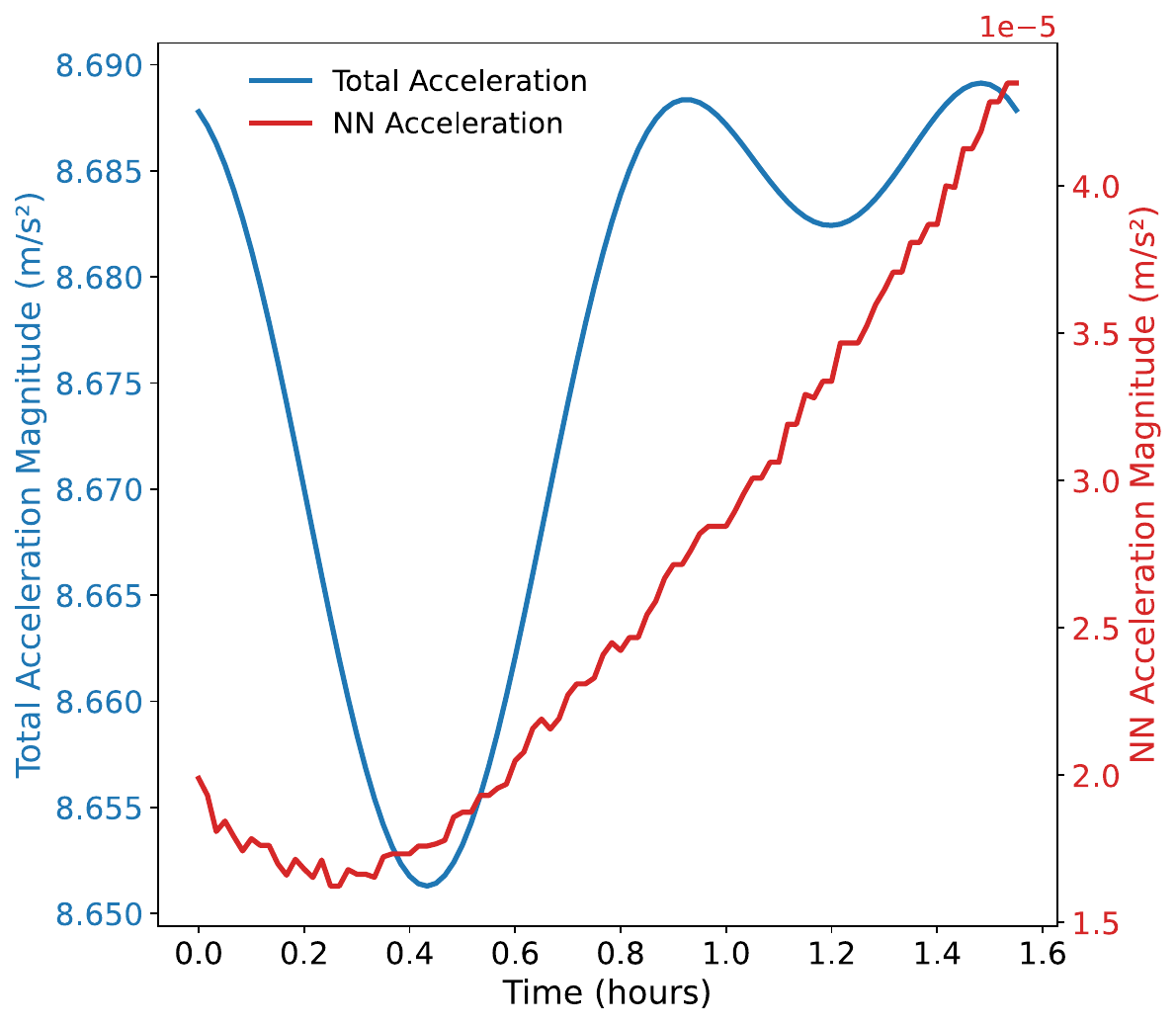}
\caption{Evolution of the learned non-conservative acceleration, in red, and the total acceleration, in blue,
for satellite~$47163$, which decayed on 2025-01-23. The figure illustrates how the Neural ODE 
separates the perturbative component from the overall dynamics during orbital decay.}
\label{fig:dual_acc_evol}
\end{figure}

An example of the temporal evolution of the learned non-conservative acceleration, 
together with the corresponding total acceleration, of a deorbiting satellite is shown in Figure~\ref{fig:dual_acc_evol}.

\begin{figure}[H]
\centering
\includegraphics[width=1\columnwidth]{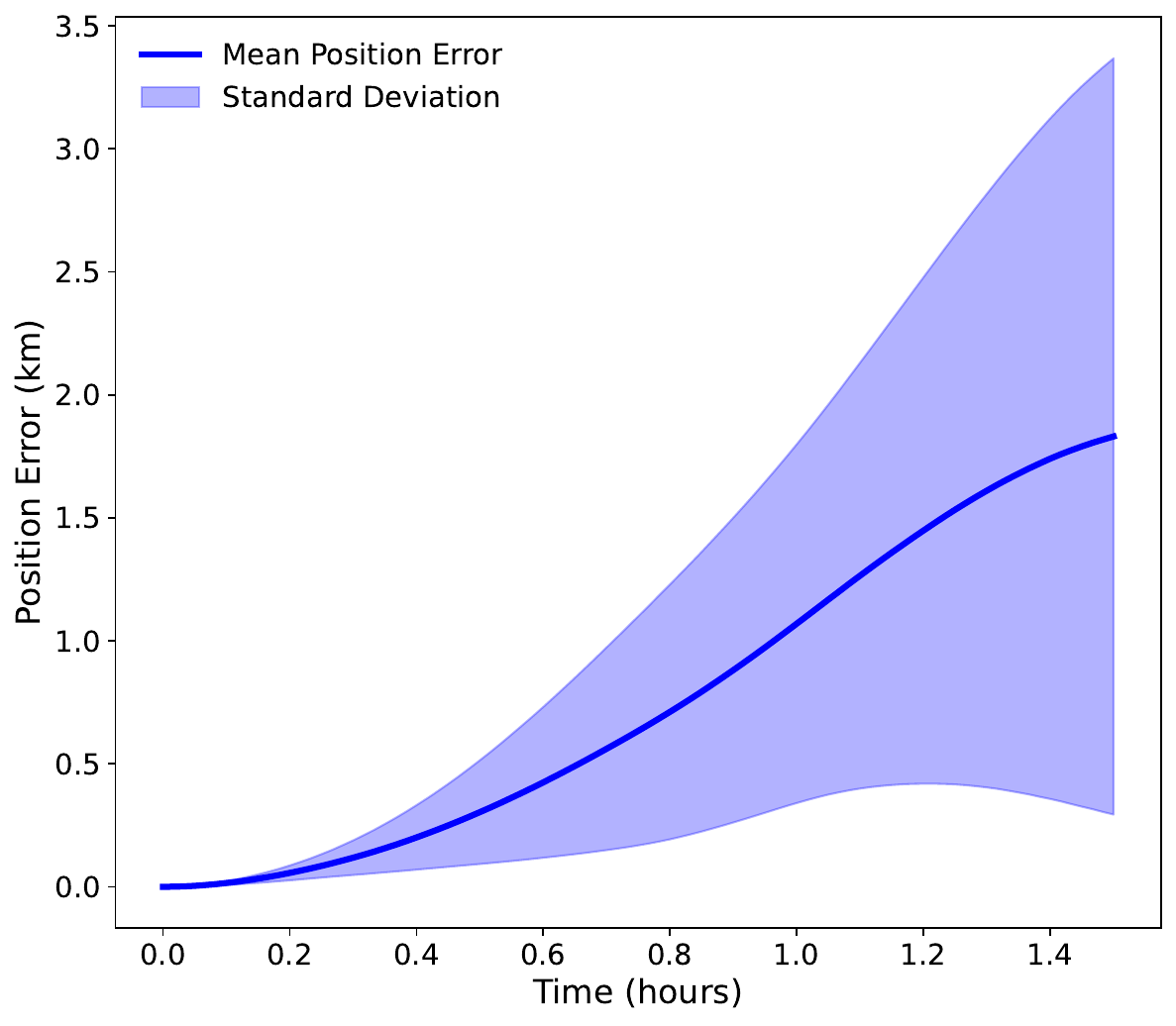}
\caption{Evolution of the mean position error over time, with the shaded region 
indicating the standard deviation across satellites.}
\label{fig:position_mean_pos}
\end{figure}

Figure~\ref{fig:position_mean_pos} shows how the mean position error evolves as a function of propagation time. As expected, the error increases with longer prediction horizons due to the accumulation of uncertainty in the orbital dynamics. However, the growth is gradual rather than exponential, and appears to plateau toward the end of the time window. This means that the orbit is not compounding small inaccuracies over the propagation. The shaded region represents the standard deviation across the 30 satellites, highlighting the variability in model performance. The widening of this band over time suggests that while some trajectories remain stable, others diverge more rapidly.
\\

\noindent In summary, the Neural ODE framework successfully captured the residual accelerations associated with non-conservative forces during orbital decay, yielding accurate predictions. The difference in learned acceleration profiles between stable and deorbiting satellites demonstrates the model's ability to adapt to different orbital regimes. While errors increased with longer propagation horizons, the growth remained sub-exponential and eventually plateaued, indicating that the dynamics were learned correctly. These findings support the use of physics informed Neural ODEs not only for robust orbital prediction but also for extracting latent forces, offering a scalable foundation for future applications.

\section{Conclusion}
\vspace{0.3cm}

\noindent This study assessed the reliability of publicly released Starlink ephemerides and introduced a Neural ODE formulation that combines known orbital physics with a learned perturbation to model residual accelerations, particularly in deorbiting regimes. Using parameters estimated per satellite, Orekit served as a high fidelity reference to benchmark 300 Starlink trajectories. This analysis showed that a non trivial fraction of the discrepancies between the overlapping forecasts of Starlink and Orekit arise from limitations within the published ephemerides themselves. Despite these limitations, the ephemerides were used to train a Neural Ordinary Differential Equation model. The framework provided more than trajectory fitting, by embedding the known physical accelerations ($-\mu \mathbf r / |\mathbf r|^3$, $J_2$, $J_3$) and restricting the network to learn only the residual perturbation, the model produced physically interpretable acceleration profiles. This design ensured physical consistency while reducing data requirements, allowing the learned component to capture non-conservative effects of deorbiting satellites.

\newpage

\section{Future Work}
\vspace{0.3cm}

\noindent While this study has demonstrated the capability of Neural ODEs to extract physically interpretable residual accelerations from Starlink ephemerides, there exist several possible avenues for future research.

\begin{itemize}
    \item The residual acceleration produced by the Neural ODE can be projected onto known physical forces (e.g., drag, solar radiation pressure, third–body effects) to infer which forces are active. Future work should focus on disentangling and quantifying these underlying contributions in the learned profiles.
    \item Extend the framework from single orbit horizons to multi orbit or multi day windows, testing the stability of the learned perturbations over longer propagation times.
    \item Explore the impact of different ODE solvers on the Neural ODE model, evaluating whether solver choice affects stability, accuracy, and the physical interpretability of the learned perturbations.
\end{itemize}

\noindent Pursuing these directions could strengthen the physical interpretability and long term stability of the learned models, while also expanding their relevance to real world scenarios in space traffic management and orbital decay forecasting. Continued development of these architectures holds potential to improve predictive performance and make them more applicable to the increasingly dynamic and congested space environment.

\section*{Acknowledgements}
This research was carried out under Project “Artificial Intelligence Fights Space Debris” Nº C626449889-0046305 co-funded by Recovery and Resilience Plan and NextGeneration EU Funds (www.recuperarportugal.gov.pt), and by NOVA LINCS (UIDB/04516/2020) with the financial support of FCT.IP.

\begin{figure}[htbp]
    \centering
    \includegraphics[width=0.15\textwidth]{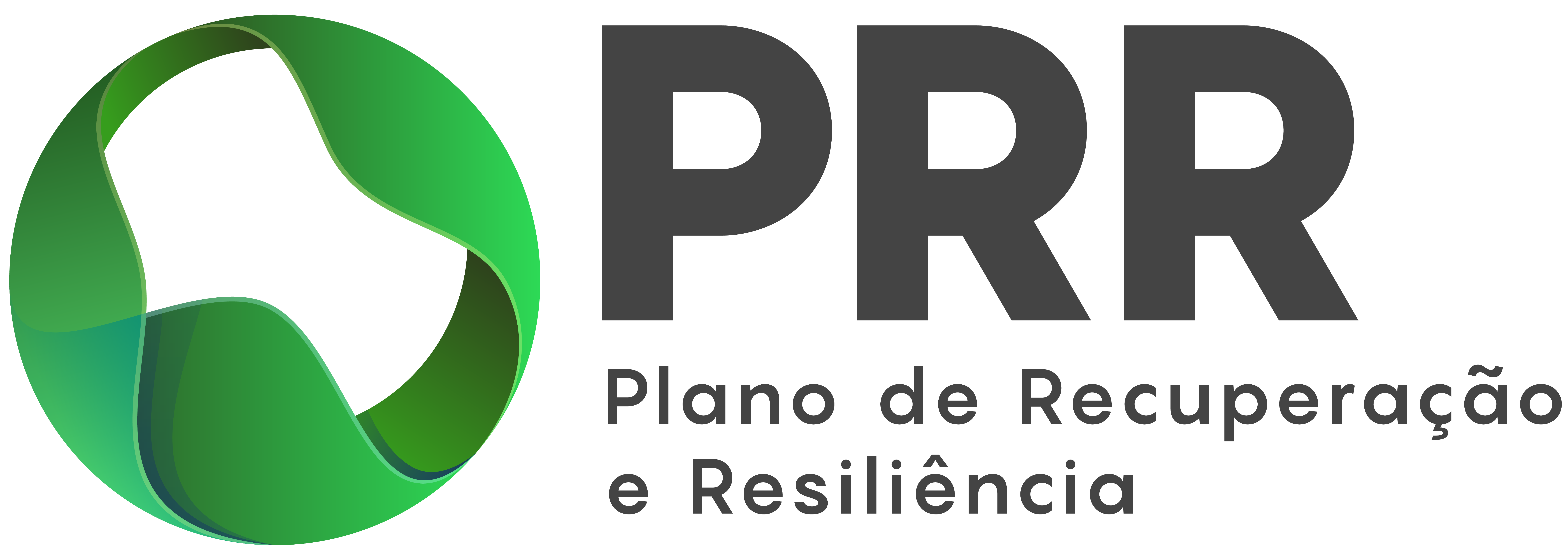} \hfill
    \includegraphics[width=0.15\textwidth]{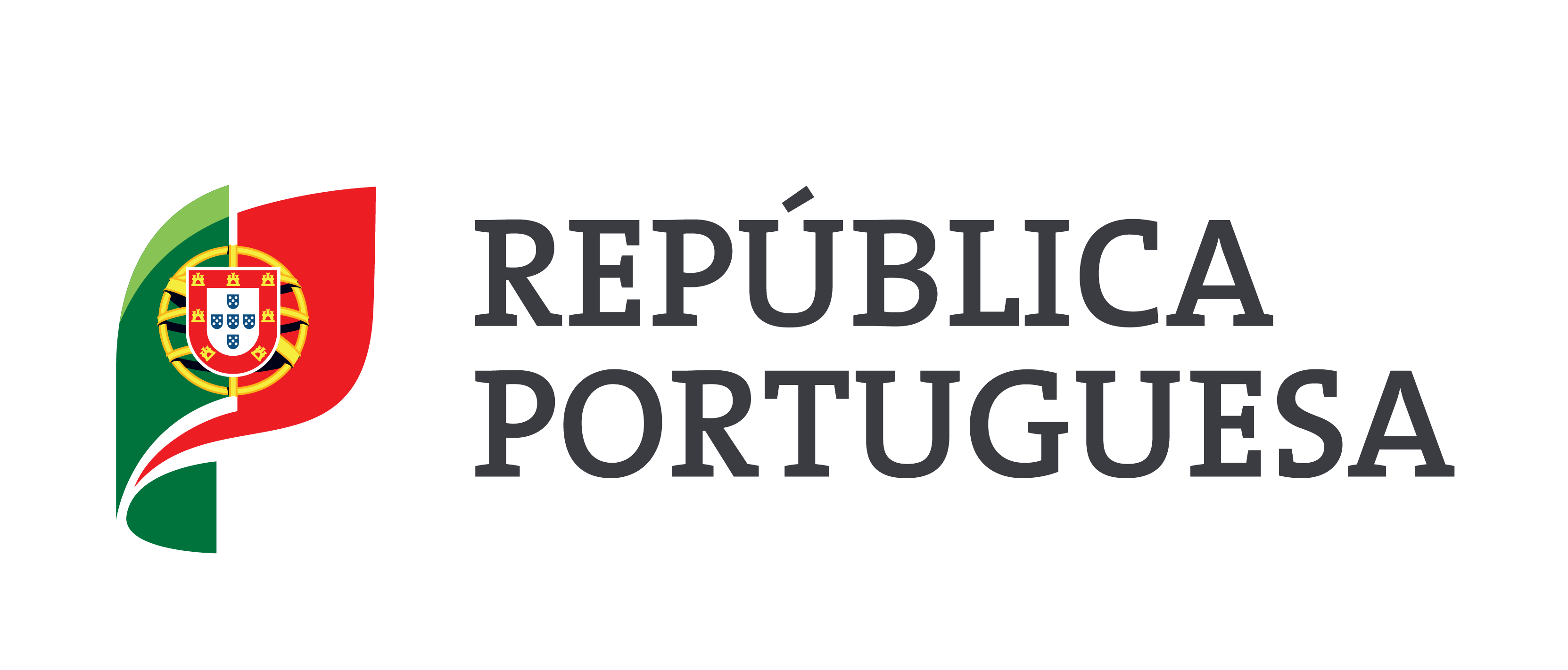} \hfill
    \includegraphics[width=0.15\textwidth]{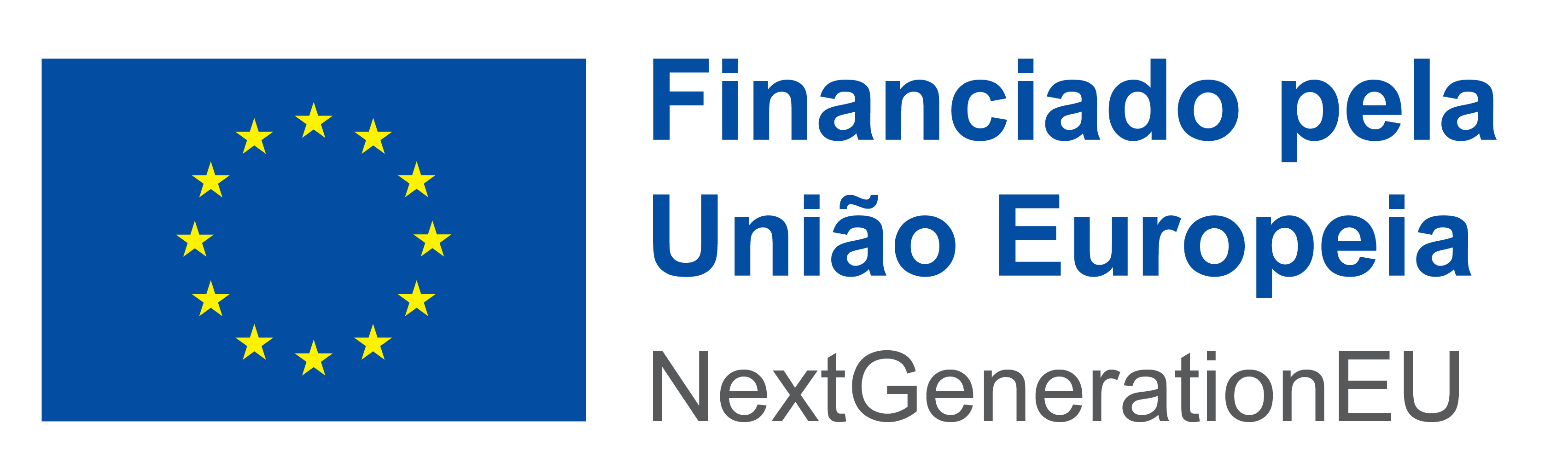}
\end{figure}

\bibliography{biblio}

%\custombibliography{biblio}  % This is the .bib file containing your bibliography data

\begin{appendices}
\end{appendices}

\clearpage

\end{document}